\newlength{\absize}
\documentclass[12pt]{article}

\usepackage{epsf}

\renewcommand{\baselinestretch}{1.5}

\begin{document}
\thispagestyle{empty}
\pagestyle{empty}
\renewcommand{\thefootnote}{\fnsymbol{footnote}}
\newcommand{\starttext}{\newpage\normalsize
\pagestyle{plain}
\setlength{\baselineskip}{3ex}\par
\setcounter{footnote}{0}
\renewcommand{\thefootnote}{\arabic{footnote}}
}

\newcommand{\preprint}[1]{\begin{flushright}
\setlength{\baselineskip}{3ex}#1\end{flushright}}
\renewcommand{\title}[1]{\begin{center}\LARGE
#1\end{center}\par}
\renewcommand{\author}[1]{\vspace{2ex}{\Large\begin{center}
\setlength{\baselineskip}{3ex}#1\par\end{center}}}
\renewcommand{\thanks}[1]{\footnote{#1}}
\renewcommand{\abstract}[1]{\vspace{2ex}\normalsize\begin{center}
\centerline{\bf Abstract}\par\vspace{2ex}\parbox{\absize}{#1
\setlength{\baselineskip}{2.5ex}\par}
\end{center}}

\newcommand{\rep}{representation}
\newcommand{\tr}{\mathop{\rm tr}}
\newcommand{\cO}{{\cal O}}
\newcommand{\cL}{{\cal L}}
\newcommand{\half}{{1\over2}}
\newcommand{\gtrsim}
{\raisebox{.2em}{$\rlap{\raisebox{-.5em}{$\;\sim$}}>\,$}}
\newcommand{\ltsim}
{\raisebox{.2em}{$\rlap{\raisebox{-.5em}{$\;\sim$}}<\,$}}
\newlength{\eqnparsize}
\setlength{\eqnparsize}{.95\textwidth}
\newcommand{\eqnbox}[1]{\parbox{\eqnparsize}{\bf\vskip.25ex
#1\vskip1ex}}
\newcommand{\PSbox}[3]
{\mbox{\rule{0in}{#3}\includegraphics{#1}\hspace{#2}}}
\def\spur#1{\mathord{\not\mathrel{#1}}}
\newcommand\etal{{\it et al.}}
\def\sech{\mathop{\rm sech}\nolimits}
\newcommand{\Ket}[1]{\left| #1 \right\rangle}

\newcommand{\svec}[1]{\stackrel{\hookrightarrow}{#1}}

\setlength{\jot}{1.5ex}
\newcommand{\figsize}{\small}
\renewcommand{\bar}{\overline}
\font\fiverm=cmr5
\input prepictex
\input pictex
\input postpictex
\input{psfig.sty}
\newdimen\tdim
\tdim=\unitlength
\def\stpltsmbl{\setplotsymbol ({\small .})}
\def\bsmbl{\setplotsymbol ({\Huge .})}
\def\tarrow{\arrow <5\tdim> [.3,.6]}
\def\barrow{\arrow <8\tdim> [.3,.6]}

%These have to do with the placing of figures
\setcounter{bottomnumber}{2}
\setcounter{topnumber}{3}
\setcounter{totalnumber}{4}
\renewcommand{\bottomfraction}{1}
\renewcommand{\topfraction}{1}
\renewcommand{\textfraction}{0}

\def\draft{
\renewcommand{\label}[1]{{\quad[\sf ##1]}}
\renewcommand{\ref}[1]{{[\sf ##1]}}
\renewenvironment{thebibliography}{\section*{References}}{}
\renewcommand{\cite}[1]{{\sf[##1]}}
\renewcommand{\bibitem}[1]{\par\noindent{\sf[##1]}}
}
% DRAFT MACROS - COMMENT THIS OUT IN FINAL VERSION
%\draft

%Move the definition below past the style to get (c.s.e) equations
\def\theequation{\thesection.\arabic{equation}}
\preprint{\#HUTP-00/A030\\ 12/00}
\title{Brane Couplings from Bulk Loops\thanks{Research supported in
part by the
National Science Foundation
under grant number NSF-PHY/98-02709.}}
\author{
Howard Georgi,\thanks{georgi@physics.harvard.edu}
Aaron K. Grant,\thanks{grant@gauss.harvard.edu}
Girma Hailu\thanks{hailu@feynman.harvard.edu}
\\
Lyman Laboratory of Physics \\
Harvard University \\
Cambridge, MA 02138
}
\date{8/00}
\abstract{ We compute loop corrections to the effective action of a field 
theory on a five-dimensional $S_1/Z_2$ orbifold.  We find that the quantum loop effects of
interactions in the bulk produce infinite contributions that require renormalization by four-dimensional couplings on
the orbifold fixed planes. Thus bulk couplings give rise to renormalization group running of brane couplings.
}

\starttext

\setcounter{equation}{0}
\section{Introduction\label{intro}}

Recently it has been proposed that large extra dimensions may be
relevant to particle physics at or near the weak scale
\cite{Arkani-Hamed:1998rs}.  This idea has opened up new possibilities
for model building that make use of extra dimensions
\cite{Pomarol:1998sd}.  In \cite{previous}, we studied a simple model
of fermions and scalars interacting on a five-dimensional space with
the fifth coordinate compactified on an $S_1/Z_2$ orbifold.  In this
model, the scalar field develops spatially varying vacuum expectation
value resulting in a ``fat brane'' structure.  The fermion field has a
chiral zero mode that can be localized near either of the orbifold
fixed points.  

In this note, we continue our analysis of the model by computing loop
corrections to the effective Lagrangian.  The orbifold boundary
conditions introduce two complications into the analysis.  First, they
break translation invariance (and hence momentum conservation) in the
fifth dimension.  Second, they single out two ``fixed points'' that
are invariant under the $Z_2$ action on $x_5$.  As a result, couplings
in the five-dimensional bulk can give rise to infinite effects that
must be renormalized by couplings on the four-dimensional orbifold
fixed planes.  This renormalization is associated with running of the
four-dimensional couplings on the fixed planes.  In the following
sections we develop the necessary formalism for computing perturbative
corrections to the effective Lagrangian, and give examples of its use
by computing the leading-logarithmic ``brane terms'' associated with
renormalization group running for several special cases.  A previous
study of perturbative field theory on orbifolds can be found in
\cite{Mirabelli:1998aj}.  This work considered a model with
supersymmetric field theories living on the fixed planes, and
discussed mechanisms for communicating supersymmetry breaking from one
brane to the other.

In section \ref{propagators}, we write down the propagators on the
orbifold. In sections \ref{fermions} and \ref{scalars}, we discuss
loop corrections. Section \ref{conclusions} contains conclusions and
some ideas for further work.

\section{Propagators\label{propagators}}

We consider a five dimensional Yukawa theory with the bulk action
\begin{equation}
\int d^5 x \,\biggl\{
\bar\psi\bigl(i\spur\,\partial-\gamma_5\partial_5 - f \phi \bigr)\psi
+ (\partial \phi)^2 - V(\phi)\biggr\}.
\end{equation}
The fifth dimension is compactified on a circle of circumference $2L$
with points on opposite sides of the circle identified.  Thus, for
instance, points $-x_5$ in $-L<-x_5<0$ are identified with points $+x_5$
in $0<x_5<L$. The points $x_5=0$ and $x_5=L$ are invariant under the
$Z_2$ action, and are referred to as fixed points.  The fields are
periodic with period $2L$, and satisfy the boundary conditions
\begin{equation}
\psi(x,-x_5)=\gamma_5\psi(x,x_5)\,,\quad
\psi(x,L+x_5)=\gamma_5\psi(x,L-x_5)\,,
\label{fermionBCs}
\end{equation}
and
\begin{equation}
\phi(x,-x_5)=-\phi(x,x_5)\,,\quad
\phi(x,L+x_5)=-\phi(x,L-x_5)\,.
\label{scalarBCs}
\end{equation}
It was shown in \cite{previous} that this model possesses a single
chiral fermion zero mode.  In addition, for suitable $V(\phi)$, the
scalar acquires a spatially varying vacuum expectation value (VEV)
$\langle\phi(x_5)\rangle$.  This spatially varying VEV can localize
the chiral zero mode near either end of the orbifold.

Now consider the propagators in this model.  If we ignore the boundary
conditions, the fermion propagator is simply that of a massless
five-dimensional fermion:
\begin{equation}
{i\over \spur p+i\gamma_5p_5}
\label{1}
\end{equation}
There are two differences on the orbifold. One is that there are true
periodic boundary conditions when $x_5\rightarrow x_5+2L$. This implies
that
\begin{equation}
p_5={\pi n\over L}
\label{2}
\end{equation}
for integer $n$.  The other difference is that because the physical region
in the orbifold is smaller than the periodicity, momentum in the $x_5$
direction is not conserved. This is related to the reflection constraints
at the orbifold boundary. An easy way to find the momentum space propagator
is to notice that we can write $\psi$ in terms of an unconstrained field
$\chi$ as
\begin{equation}
\psi(x,x_5)=\half\,\Bigl(\chi(x,x_5)+\gamma_5\,\chi(x,-x_5)\Bigr)\,.
\label{unconstrained}
\end{equation}
This field automatically satisfies \ref{fermionBCs}.  We can now use this
to compute the momentum space propagator. Notice that since both $x_5$ and
$-x_5$ appear in (\ref{unconstrained}), the propagator
\begin{equation}
S_5(x-x',x_5,x_5')=\left\langle \psi(x,x_5)\,\bar\psi(x',x_5')\right\rangle
\label{5}
\end{equation}
depends on both $x_5-x_5'$ and $x_5+x_5'$. Doing the Fourier transform gives
\begin{equation}
{i\over4}\left\{
{\delta_{p_5^{}p_5'}\over \spur p+i\gamma_5p_5}
+\gamma_5{\delta_{-p_5^{}p_5'}\over \spur p-i\gamma_5p_5}
-{\delta_{-p_5^{}p_5'}\over \spur p+i\gamma_5p_5}\gamma_5
-\gamma_5{\delta_{p_5^{}p_5'}\over \spur p-i\gamma_5p_5}\gamma_5
\right\}\,.
\label{6}
\end{equation}
This can be simplified to
\begin{equation}
{i\over2}\left\{
{\delta_{p_5^{}p_5'}\over \spur p+i\gamma_5p_5}
-{\delta_{-p_5^{}p_5'}\over \spur p+i\gamma_5p_5}\gamma_5
\right\}\,.
\label{7}
\end{equation}

Similarly, we can find the scalar propagator by rewriting $\phi$ in terms
of an unconstrained field $\Phi$ as
\begin{equation}
\phi(x,x_5) = \half\,\Bigl(\Phi(x,x_5)-\Phi(x,-x_5)\Bigr)\,.
\end{equation}
This gives a propagator
\begin{equation}
{i\over2}\left\{
{\delta_{p_5^{}p_5'}-\delta_{-p_5^{}p_5'}\over p^2-p_5^2}
\right\}\,.
\label{9}
\end{equation}

\setcounter{equation}{0}
\section{Fermions\label{fermions}}

Now consider the one-loop correction to the fermion propagator from
the diagram in fig.\ref{fig-1}.  The fermion has momentum $(p,p_5')$
coming in and momentum $(p,p_5)$ going out. Momentum is conserved at
the vertices. So say that the incoming fermion splits into a fermion
with momentum $(k,k_5')$ and a scalar with momentum
$(p-k,p_5'-k_5')$. These propagate and the 5 components change drop
their primes and get reabsorbed. The internal loop momentum $k$ is
integrated and $k_5$ and $k_5'$ are summed over. The diagram is then
\begin{equation}
i\Sigma = \frac{f^2}{4}\sum_{k_5^{},k_5'}\int \frac{d^D k}{(2\pi)^D}\,\left\{
{\delta_{k_5^{}k_5'}\over \spur k+i\gamma_5k_5}
-{\delta_{-k_5^{}k_5'}\over \spur k+i\gamma_5k_5}\gamma_5
\right\}
\left\{
{\delta_{(p_5^{}-k_5^{}),(p_5'-k_5')}
-\delta_{-(p_5^{}-k_5^{}),(p_5'-k_5')}\over (p-k)^2-(p_5-k_5)^2}
\right\}\,.
\label{FermionGraph}
\end{equation}
{\figsize\begin{figure}[htb]
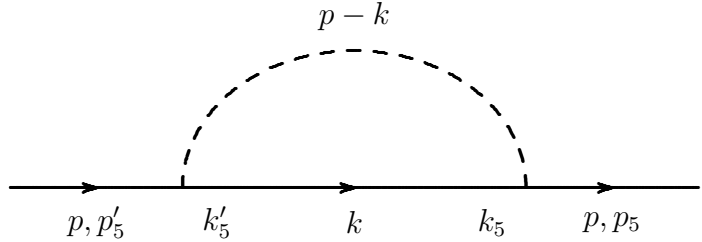

$$\beginpicture
\setcoordinatesystem units <1.3\tdim,1.3\tdim>
\stpltsmbl
\tarrow from -100 0 to -75 0
\tarrow from -50 0 to 0 0
\tarrow from 50 0 to 75 0
\plot -100 0 100 0 /
\put {$k$} at 0 -10
\put {$k_5'$} at -40 -10
\put {$k_5$} at 40 -10
\put {$p,p_5'$} at -75 -10
\put {$p,p_5$} at 75 -10
\put {$p-k$} at 0 50
\setdashes
\ellipticalarc axes ratio 5:4 180 degrees from 50 0 center at 0 0
\linethickness=0pt
\putrule from 0 60 to 0 -20
\putrule from -100 0 to 100 0
\endpicture$$
\caption{\figsize\sf\label{fig-1}One loop correction to the 
fermion propagator.}\end{figure}}
Summing over $k_5'$, the integrand becomes
\begin{equation}
{
1
\over
(\spur k+i\gamma_5k_5)[ (p-k)^2-(p_5-k_5)^2]}
\left\{ \delta_{p_5^{}p_5'} + \delta_{p_5^{}-p_5'} \gamma_5
       - \delta_{2 k_5^{} , ( p_5 + p_5' )} - \delta_{2 k_5^{}, ( p_5-p_5')}
         \gamma_5
\right\}
\label{FermionIntegrand}
\end{equation}
\begin{equation}
={
\spur k+i\gamma_5k_5
\over
(k^2-k_5^2)[ (p-k)^2-(p_5-k_5)^2]}
\left\{ \delta_{p_5^{}p_5'} + \delta_{p_5^{}-p_5'} \gamma_5
       - \delta_{2 k_5^{} , ( p_5 + p_5' )} - \delta_{2 k_5^{}, ( p_5-p_5')}
         \gamma_5
\right\}
\label{FermionIntegrand2}
\end{equation}

When we do the $D$-dimensional integral in (\ref{FermionGraph}), we
encounter $1/\epsilon$ pole terms (where $D=4-2 \epsilon$). In this
paper, we consider only these divergent terms. For the pole terms, the
$p$ dependence comes only from the $\delta$-functions and the
numerator in (\ref{FermionIntegrand2}).  The first two terms in braces
in (\ref{FermionIntegrand2}) give contributions where $|p_5|$ is
conserved.  These terms are contributions to the five-dimensional bulk
fermion kinetic energy.  However, the last two terms have a different
structure.  They do not conserve $|p_5|$ and therefore cannot be
associated with any term in the bulk Lagrangian. Rather, they yield a
sum of terms where $p_5\pm p_5'$ changes by an even multiple of
$\pi/L$.  These terms give contributions to the action that depend
only on the values of the fields at the orbifold fixed points
$x_5=0,L$, and thus they renormalize the couplings on the brane.  We
can understand this by considering a generic momentum space operator
like
\begin{equation}
\sum_{p_5 = p_5' + 2 \pi n / L } \bar{\psi}(p,p_5) \Gamma \psi(p,p_5'),
\end{equation}
where $\Gamma$ is some Dirac matrix.  Transforming this to position
space gives
\begin{equation}
\left(\delta(x_5) + \delta(L-x_5) \right) \bar{\psi}(x,x_5) \Gamma \psi(x,x_5).
\end{equation}
The constraint that $p_5$ changes by an {\it even} multiple of $\pi / L$
means that we get $\delta$-functions at $x_5=0,\pm L,\pm 2 L,\dots$.  We
have explictly written the $\delta$-functions that are singular in the
physical region $0\leq x_5 \leq L$.  If all multiples of $\pi/L$ were
summed over, we would of course get $\delta$-functions at $x_5=0,\pm 2L,
\pm 4L,\dots$.

Explicitly evaluating (\ref{FermionGraph}), we encounter a divergent piece
\begin{equation}
i \Sigma = \frac{-i}{4} \frac{f^2}{16\pi^2} 
\biggl[ \spur{p} \left( \frac{1+\gamma_5}{2} \right)
+ i p_5  \left( \frac{1+\gamma_5}{2} \right)
- i p_5' \left( \frac{1-\gamma_5}{2} \right)
\biggr]\frac{1}{\epsilon} + \dots\,.
\end{equation}
When we eliminate the pole by minimal subtraction, we are
renormalizing a brane term. This contributes to the running of the
corresponding term on the brane.  Subtracting and converting back to
position space gives the contribution to the effective Lagrangian:
\begin{equation}
\delta L_{\rm eff}(\mu) = \frac{1}{2} \frac{f^2}{16\pi^2}
\log\left(\frac{\mu}{M}\right)
\left[ \delta(x_5) + \delta(x_5-L) \right]
\left[
\bar\psi_+ i\spur{\partial} \psi_+ 
+(\partial_5 \bar\psi_-) \psi_+ 
+\bar\psi_+ (\partial_5 \psi_-) 
\right]
\label{OneLoopFermionAction}
\end{equation}
where $\psi_\pm = (1/2) ( 1\pm \gamma_5 ) \psi$.

% The results we have obtained are consistent with the plausible notion
% that there is something special about the orbifold in which all the
% fields are smooth at and across the orbifold fixed point. Nontrivial
% physics on the brane could spoil this smoothing.  However, the
% particular combination (\ref{OneLoopFermionAction}) induced by the
% renormalization group does not affect the modes that we obtaining by
% ignoring the physics on the brane and assuming smoothness. The
% functional derivative of (\ref{OneLoopFermionAction}) with respect to
% $\bar\psi_+$,
% \begin{equation}
% -(i\spur\partial\,\psi_++\partial_5\psi_-)\,,
% \label{26}
% \end{equation}
% already vanishes because of the equations of motion in the bulk
% evaluated at $x_5=0$ or $L$ where $\phi=0$. The functional derivative
% with respect to $\bar\psi_-$ vanishes because of the boundary
% conditions.

\setcounter{equation}{0}
\section{Scalars\label{scalars}}

In this section, we consider the divergent contributions to loops
involving external scalars.  The one-loop scalar tadpole is shown in
Fig.~\ref{fig-3}.  {\figsize\begin{figure}[htb]
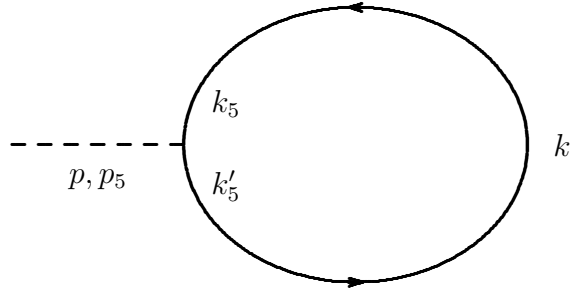

$$\beginpicture
\setcoordinatesystem units <1.3\tdim,1.3\tdim>
\stpltsmbl
\tarrow from -2 -40 to 2 -40
\tarrow from 2 40 to -2 40
\setdashes
\plot -100 0 -50 0 /
\setsolid
\put {$k$} at 60 0
\put {$k_5'$} [l] at -42 -12
\put {$k_5$} [l] at -42 12
\put {$p,p_5$} at -75 -10
\ellipticalarc axes ratio 5:4 360 degrees from 50 0 center at 0 0
\linethickness=0pt
\putrule from 0 60 to 0 -60
\putrule from -100 0 to 60 0
\endpicture$$
\caption{\figsize\sf\label{fig-3}One loop contribution to the scalar
tadpole.}
\end{figure}}
This diagram yields
\begin{equation}
\frac{f}{2} \sum_{k_5} \int\, \frac{d^D k}{(2\pi)^D} {\rm Tr}\,
\frac{ ( \spur k + i \gamma_5 k_5 )
( \delta_{k_5,k_5+p_5} - \gamma_5 \delta_{k_5,-(k_5+p_5)})}
{  k^2 - k_5^2  },
\end{equation}
where we have used momentum conservation at the vertex to write $k_5'
= k_5+p_5$.  As before, the first Kronecker-$\delta$ has the form of a
renormalization of the bulk Lagrangian (the coefficient vanishes in
this case), while the second yields a brane term.  Evaluating the
integral with dimensional regularization and minimal subtraction gives
\begin{equation}
\delta L = \frac{f}{32\pi^2}\log\left(\frac{\mu}{M}\right)
(\delta(x_5) + \delta(x_5-L) ) \partial_5^3 \phi.
\end{equation}
In cutoff regularization, we would also find a quadratic divergence
proportional to 
\begin{equation}
(\delta(x_5)+\delta(L-x_5)) \partial_5 \phi\,.
%\label{}
\end{equation}
The
effect of the $DR\bar{MS}$ term can be made more tranparent by a
change of variables in $\phi$.  For instance if the scalar potential
vanishes, then we can eliminate the tadpole from the scalar sector of
the theory by making the substitution
\begin{equation}
\phi = \phi' - \frac{f}{32\pi^2} \log\left(\frac{\mu}{M}\right)
( \delta'(x_5) + \delta'(x_5-L) )\,.
\end{equation}
This shift introduces a term proportional to $(\delta'(x_5) +
\delta'(x_5-L))$ fermion equation of motion.
{\figsize\begin{figure}[htb]
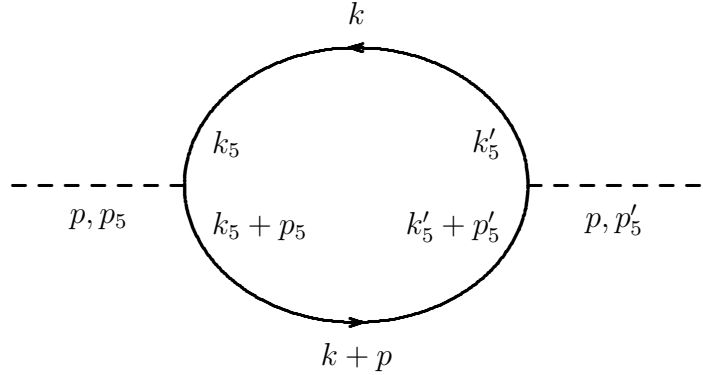

$$\beginpicture
\setcoordinatesystem units <1.3\tdim,1.3\tdim>
\stpltsmbl
\tarrow from -2 -40 to 2 -40
\tarrow from 2 40 to -2 40
\setdashes
\plot -100 0 -50 0 /
\plot 50 0 100 0 /
\setsolid
\put {$k+p$} at 0 -50
\put {$k_5+p_5$} [l] at -42 -12
\put {$k_5'+p_5'$} [r] at 42 -12
\put {$k_5$} [l] at -42 12
\put {$k_5'$} [r] at 42 12
\put {$p,p_5$} at -75 -10
\put {$p,p_5'$} at 75 -10
\put {$k$} at 0 50
\ellipticalarc axes ratio 5:4 360 degrees from 50 0 center at 0 0
\linethickness=0pt
\putrule from 0 60 to 0 -60
\putrule from -100 0 to 100 0
\endpicture$$
\caption{\figsize\sf\label{fig-2}One loop correction to the scalar
propagator.}
\end{figure}}

The one loop contribution to the scalar propagator from the diagram in
figure~\ref{fig-2} gives no contribution to interactions on the brane.  In
this case, the loop integral is
\begin{eqnarray}
&&- \frac{f^2}{4}\sum_{k_5^{},k_5'}\int \frac{d^D k}{(2\pi)^D}\,
{\rm Tr}\,
\frac{ ( \spur k + i\gamma_5 k_5 ) 
( \delta_{k_5,k_5'} - \gamma_5 \delta_{k_5,-k_5'} )}{k^2 -
k_5^2}\nonumber\\
&\times&
\frac{ ( \spur k +\spur p + i\gamma_5 [ k_5' + p_5' ] ) 
( \delta_{k_5+p_5,k_5'+p_5'} - \gamma_5 \delta_{k_5+p_5,-k_5'-p_5'} )}
{(k+p)^2 - (k_5'+p_5')^2}.
\end{eqnarray}
The brane terms vanish, since they are proportional to traces of odd
numbers of $\gamma$ matrices, or traces of fewer than four Dirac
matrices with $\gamma_5$.  From symmetry considerations alone, one
might have expected to find brane terms proportional to $(\partial_5
\phi)^2$.  At higher loops, such terms are indeed generated.  To
investigate this, let's consider the two-loop graph in
figure~\ref{fig-4}.  {\figsize\begin{figure}[htb]
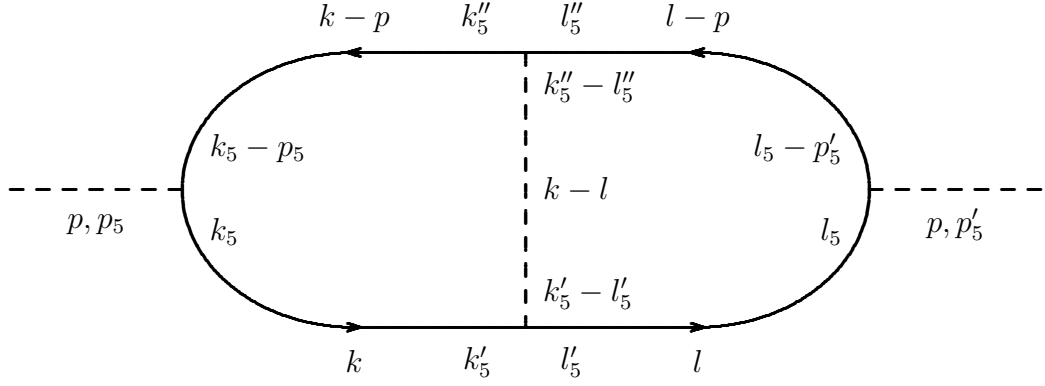

$$\beginpicture
\setcoordinatesystem units <1.3\tdim,1.3\tdim>
\stpltsmbl
\tarrow from -2 -40 to 2 -40
\tarrow from 2 40 to -2 40
\tarrow from 98 -40 to 102 -40
\tarrow from 102 40 to 98 40
\plot 0 40 100 40 /
\plot 0 -40 100 -40 /
\setdashes
\plot -100 0 -50 0 /
\plot 150 0 200 0 /
\plot 50 40 50 -40 /
\setsolid
\put {$k$} at 0 -50
\put {$l$} at 100 -50
\put {$k_5$} [l] at -42 -12
\put {$l_5$} [r] at 142 -12
\put {$k_5'$} [r] at 40 -50
\put {$l_5'$} [l] at 60 -50
\put {$k_5''$} [r] at 40 50
\put {$l_5''$} [l] at 60 50
\put {$k-l$} [l] at 55 0
\put {$k_5'-l_5'$} [l] at 55 -30
\put {$k_5''-l_5''$} [l] at 55 30
\put {$k_5-p_5$} [l] at -42 12
\put {$l_5-p_5'$} [r] at 142 12
\put {$p,p_5$} at -75 -10
\put {$p,p_5'$} at 175 -10
\put {$k-p$} at 0 50
\put {$l-p$} at 100 50
\ellipticalarc axes ratio 5:4 180 degrees from 0 40 center at 0 0
\ellipticalarc axes ratio 5:4 -180 degrees from 100 40 center at 100 0
\linethickness=0pt
\putrule from 0 60 to 0 -60
\putrule from -100 0 to 100 0
\endpicture$$
\caption{\figsize\sf\label{fig-4}Two loop correction to the scalar
propagator.}
\end{figure}}
Now consider the conservation of the 5 component of the loop momentum around
the loop. Each of the propagators conserves the 5 component of the momentum
it carries up to a factor of $\pm1$. Call these factors $\eta$s, and
associate the $\eta$s with propagators as shown in
figure~\ref{fig-42}. Then we have
\begin{equation}
k_5-p_5=\eta_1k_5''\,,\quad
l_5-p'_5=\eta_2l_5''\,,\quad
k_5=\eta_3k_5'\,,\quad
l_5=\eta_4l_5'\,,\quad
k_5''-l_5''=\eta_5(k_5'-l_5')\,.
\label{30}
\end{equation}
{\figsize\begin{figure}[htb]
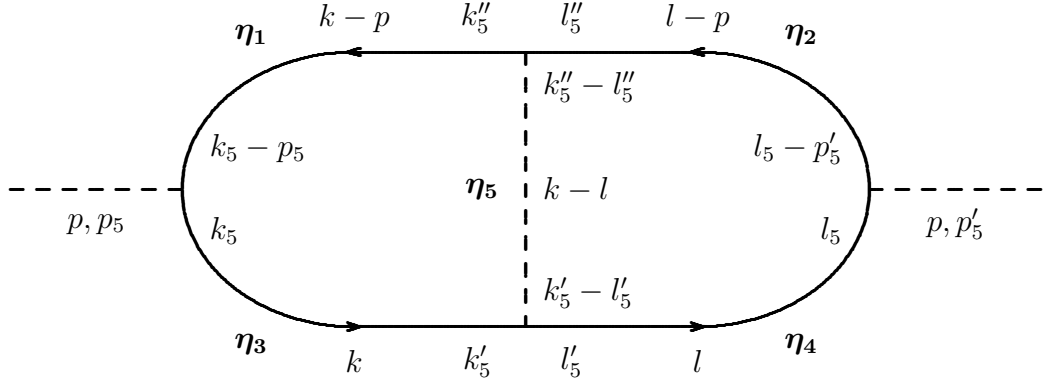

$$\beginpicture
\setcoordinatesystem units <1.3\tdim,1.3\tdim>
\stpltsmbl
\tarrow from -2 -40 to 2 -40
\tarrow from 2 40 to -2 40
\tarrow from 98 -40 to 102 -40
\tarrow from 102 40 to 98 40
\plot 0 40 100 40 /
\plot 0 -40 100 -40 /
\setdashes
\plot -100 0 -50 0 /
\plot 150 0 200 0 /
\plot 50 40 50 -40 /
\setsolid
\put {$k$} at 0 -50
\put {\boldmath$\eta_3$} at -30 -45
\put {\boldmath$\eta_4$} at 130 -45
\put {$l$} at 100 -50
\put {$k_5$} [l] at -42 -12
\put {$l_5$} [r] at 142 -12
\put {$k_5'$} [r] at 40 -50
\put {$l_5'$} [l] at 60 -50
\put {$k_5''$} [r] at 40 50
\put {$l_5''$} [l] at 60 50
\put {$k-l$} [l] at 55 0
\put {\boldmath$\eta_5$} [r] at 42 0
\put {$k_5'-l_5'$} [l] at 55 -30
\put {$k_5''-l_5''$} [l] at 55 30
\put {$k_5-p_5$} [l] at -42 12
\put {$l_5-p_5'$} [r] at 142 12
\put {$p,p_5$} at -75 -10
\put {$p,p_5'$} at 175 -10
\put {$k-p$} at 0 50
\put {\boldmath$\eta_1$} at -30 45
\put {\boldmath$\eta_2$} at 130 45
\put {$l-p$} at 100 50
\ellipticalarc axes ratio 5:4 180 degrees from 0 40 center at 0 0
\ellipticalarc axes ratio 5:4 -180 degrees from 100 40 center at 100 0
\linethickness=0pt
\putrule from 0 60 to 0 -60
\putrule from -100 0 to 100 0
\endpicture$$
\caption{\figsize\sf\label{fig-42} The association of the $\eta$s with the 
propagators in figure~\ref{fig-4}.}
\end{figure}}
Eliminating $k_5'$, $l_5'$, $k_5''$ and $l_5''$ from (\ref{30}) gives
\begin{equation}
\eta_1p_5-\eta_2p_5'=(\eta_1-\eta_3\eta_5)k_5
+(\eta_2-\eta_4\eta_5)l_5\,.
\label{31}
\end{equation}
We get brane terms when the right hand side of (\ref{31}) does not
vanish. It vanishes only when
$\eta_1\eta_3\eta_5=\eta_2\eta_4\eta_5=1$. This result is easy to
remember. $\eta_1$, $\eta_3$ and $\eta_5$ are the $\eta$s associated with
the $k$ loop and $\eta_2$, $\eta_4$ and $\eta_5$ are the $\eta$s associated
with the $l$ loop. The product in each loop must be $+1$ to give a bulk
term. Otherwise we get a brane term. This works for the one-loop diagrams
as well, so we may speculate that it is a general result. 

Of course, there is a second issue, namely whether or not the Dirac trace
in the self-energy diagram vanishes.  This gives a second constraint on the
$\eta$'s.  It is easy to see that we must have
$\eta_1 \eta_2 \eta_3 \eta_4 = +1$ to get a non-zero result. 

It is clear that from diagrams like figure~\ref{fig-4} we can get brane
contributions with an even number of $\partial_5$s. Perhaps we should not
read too much into their absence at the one-loop level.

Next consider the one-loop contribution to the scalar three point function
in figure~\ref{fig-5}.
{\figsize\begin{figure}[htb]
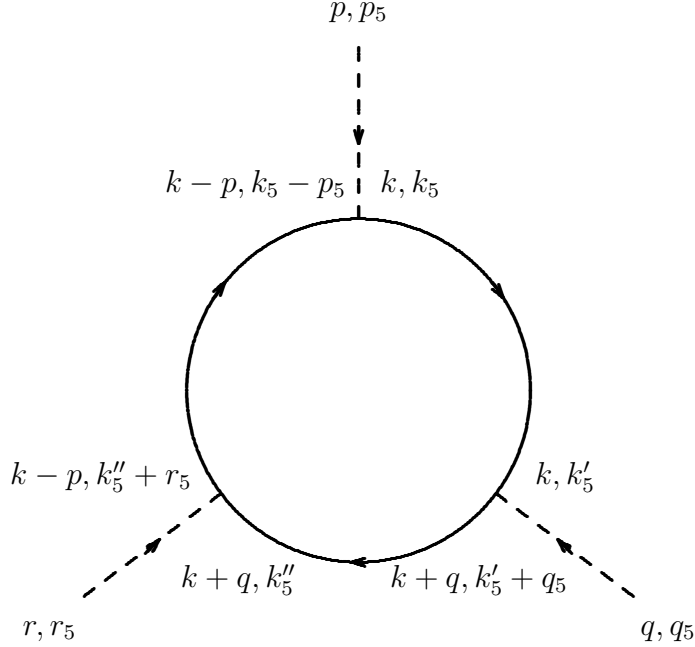

$$\beginpicture
\setcoordinatesystem units <1.3\tdim,1.3\tdim>
\stpltsmbl
\tarrow from 2 -50 to -2 -50
\tarrow from 39 31.5 to 42 27
\tarrow from -42 27 to  -39 31.5
\setdashes
\plot 0 50 0 100 /
\plot 40 -30 80 -60 /
\plot -40 -30 -80 -60 /
\setsolid
\tarrow from 0 77 to 0 72
\tarrow from 62 -47.5 to 58 -43.5 
\tarrow from -62 -47.5 to -58 -43.5 
%\put {$k$} at 0 -60 
\circulararc 360 degrees from 50 0 center at 0 0
\linethickness=0pt
\putrule from 0 110 to 0 -70
\putrule from -100 0 to 100 0
\put {$q,q_5$} at 90 -70
\put {$p,p_5$} at 0 110
\put {$r,r_5$} at -90 -70
\put {$k,k_5$} at 15 60
\put {$k-p,k_5-p_5$} at -30 60
\put {$k-p,k_5''+r_5$} at -75 -25
\put {$k+q,k_5''$} at -35 -55
\put {$k+q,k_5'+q_5$} at 35 -55
\put {$k,k_5'$} at 60 -25
\endpicture$$
\caption{\figsize\sf\label{fig-5}One loop contribution to the scalar
3-point  function.}
\end{figure}}
The contribution to the effective Lagrangian is constrained by the
boundary conditions.  A $\phi^3$ term is inconsistent with the
boundary conditions, whether it is in the bulk or on the brane.  Terms
like $\phi^2 \partial_5 \phi$ are consistent with the boundary
conditions, but vanish on the brane.  The lowest dimension operator
that is non-zero on the brane and consistent with the boundary
conditions is $(\partial_5 \phi)^3$.  Now consider the loop integral.
Labelling the momenta as shown in figure~\ref{fig-5}, the loop
integral is
\begin{eqnarray}
&&\frac{f^3}{8}\sum_{k_5,k_5',k_5''}\int \frac{d^D k}{(2\pi)^D}
{\rm Tr}\,
\frac{ 
( \spur k + i \gamma_5 k_5' )
	( \delta_{k_5,k_5'}-\gamma_5 \delta_{k_5,-k_5'} )}
{k^2-k_5'^2}\nonumber\\
&\times&\frac{( \spur k -\spur p + i \gamma_5 ( k_5 - p_5) )
	( \delta_{k_5-p_5,k_5''+r_5}-\gamma_5 \delta_{k_5-p_5,-k_5''-r_5} )}
{[k+q]^2-[k_5-p_5]^2}\nonumber\\
&\times&{\frac{( \spur k +\spur q + i \gamma_5 k_5'') )
	( \delta_{k_5'',k_5'+q_5}-\gamma_5 \delta_{k_5'',-k_5'-q_5} )}
{[k-p]^2-k_5''^2}}
\end{eqnarray}
As in the case of the self-energy diagram, the brane terms come from
cross terms where the 5 component of the loop momentum undergoes an
odd number of sign changes as it flows around the loop.  Contributions
to the running come from brane terms with two or three powers of the
loop momentum in the numerator.  We can see that there is no $\phi^3$
term on the brane: this is simply because the portions of the
integrand that would yield such a term are proportional to traces of
the form
\begin{equation}
{\rm Tr}\, \gamma_\mu \gamma_\nu \gamma_\lambda \gamma_5 = 0.
\end{equation}
We can also see that no term of the form $\phi^2 \partial_5\phi$ is
induced.  Such a term would vanish on the brane, but is nonetheless
consistent with the boundary conditions.  Collecting all terms in the
numerator that are linear in the 5 components of the external momenta,
we find a complete cancellation. This means that there are no terms of
the form $\phi^2 \partial_5\phi$, whether finite or infinite.  We
expect that the one-loop correction will, however, generate finite
corrections with three or more derivatives.

Now consider the one loop correction to the $\phi\bar\psi\psi$
coupling shown in fig.~\ref{fig-6}.  
{\figsize\begin{figure}[htb]
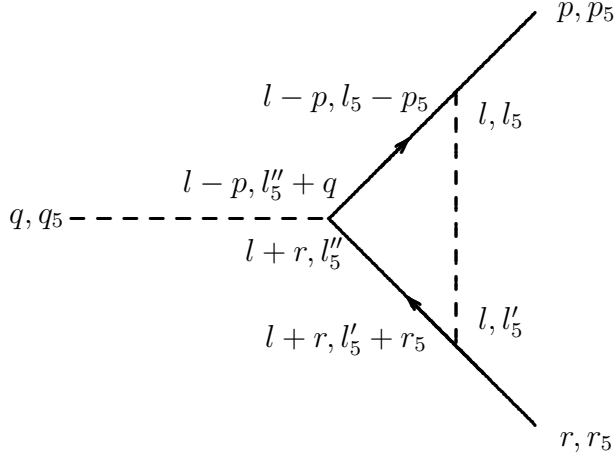

$$\beginpicture
\setcoordinatesystem units <1.3\tdim,1.3\tdim>
\stpltsmbl
\tarrow from 15 15 to 23 23
\tarrow from 45 -45 to 23 -23
\plot 0 0 60 60 /
\plot 0 0 60 -60 /
\setdashes
\plot 37 37 37 -37 /
\plot -75 0 0 0 /
\put {$q,q_5$} at -85 0
\put {$p,p_5$} at 75 60
\put {$r,r_5$} at 75 -65
\put {$l,l_5$} at 50 30
\put {$l,l_5'$} at 50 -30
\put {$l+r,l_5'+r_5$} at 5 -35
\put {$l+r,l_5''$} at -10 -10
\put {$l-p,l_5''+q$} at -20 10
\put {$l-p,l_5-p_5$} at 5 35
\linethickness=0pt
\putrule from -75 0 to 75 0 
\putrule from -75 -60 to -75 60
\endpicture$$
\caption{\figsize\sf\label{fig-6}One loop correction to the 
Yukawa coupling.}\end{figure}} It's easy to see that the divergent
part of the brane term vanishes.  The divergence would come from terms
with two powers of the loop momentum in the numerator of the integrand.  A
short computation shows that this piece of the numerator is
proportional to
\begin{equation}
\spur \,l \spur \,l+
\spur \,l \spur \,l\gamma_5+
\spur \,l \gamma_5 \spur \,l+
\spur \,l \gamma_5 \spur \,l \gamma_5 = 0.
\end{equation}
Hence there are no infinite renormalizations of the $\phi \psi
\bar\psi$ coupling on the brane.

\setcounter{equation}{0}
\section{Conclusions\label{conclusions}}

Field theories on orbifolds may be a useful tool for model building in
extra dimensions.  We have shown that these theories necessarily have
a hybrid structure, involving both five-dimensional bulk couplings and
four-dimensional brane couplings.  Under renormalization group flow, a
theory with no brane couplings will generally flow to a theory with
non-trivial physics on the brane. It is important to note that what we
have discussed in this paper is the renormalization group running of
couplings in the five-dimensional theory. Both the bulk and the brane
couplings are defined in the five-dimensional theory, although by
definition, the brane couplings appear in the Lagrangian with a
$\delta$-function that restricts them to the brane. This does not
directly tell us about the running in the couplings in an effective
four-dimensional theory derived from the five-dimensional physics,
although it is surely a necessary component of any consistent
calculation of this running. We hope to return to this issue and to
study the particle physics implications of this result in future work.

\section*{Acknowledgements}

Some of this work was done at the Aspen Center for Physics during the
workshop on Physics at the Weak Scale.  HG is grateful to the center for
providing a splendid work environment and to many participants in the
workshop for useful converstions, particularly Gia Dvali, Lawrence Hall,
Lisa Randall, Martin Schmaltz and Misha Voloshin.


\begin{thebibliography}{99}
\frenchspacing

\bibitem{Arkani-Hamed:1998rs}
See for example
N.~Arkani-Hamed, S.~Dimopoulos and G.~Dvali,
%``The hierarchy problem and new dimensions at a millimeter,''
Phys.\ Lett.\  {\bf B429}, 263 (1998)
[hep-ph/9803315];
%%CITATION = HEP-PH 9803315;%%
%\href{\wwwspires?eprint=HEP-PH/9803315}{SPIRES}
L.~Randall and R.~Sundrum,
%``A large mass hierarchy from a small extra dimension,''
Phys.\ Rev.\ Lett.\  {\bf 83}, 3370 (1999)
[hep-ph/9905221].
%%CITATION = HEP-PH 9905221%%


%\cite{Pomarol:1998sd}
\bibitem{Pomarol:1998sd}
See for example A.~Pomarol and M.~Quiros,
%``The standard model from extra dimensions,''
Phys.\ Lett.\  {\bf B438}, 255 (1998)
[hep-ph/9806263];
%%CITATION = HEP-PH 9806263;%%
%\href{\wwwspires?eprint=HEP-PH/9806263}{SPIRES}
%\cite{Arkani-Hamed:2000dc}:
N.~Arkani-Hamed and M.~Schmaltz,
%``Hierarchies without symmetries from extra dimensions,''
Phys.\ Rev.\  {\bf D61}, 033005 (2000)
[hep-ph/9903417].
%%CITATION = HEP-PH 9903417;%%

%previous paper
\bibitem{previous}
%\cite{Georgi:2000wb}
%\bibitem{Georgi:2000wb}
H.~Georgi, A.~Grant and G.~Hailu,
%``Chiral fermions, orbifolds, scalars and fat branes,''
hep-ph/0007350.
%%CITATION = HEP-PH 0007350;%%%previous paper

%\cite{Mirabelli:1998aj}:
\bibitem{Mirabelli:1998aj}
E.~A.~Mirabelli and M.~E.~Peskin,
%``Transmission of supersymmetry breaking from a 4-dimensional boundary,''
Phys.\ Rev.\  {\bf D58}, 065002 (1998)
[hep-th/9712214].
%%CITATION = HEP-TH 9712214;%%



\end{thebibliography}
\end{document}